\documentclass[10pt,twocolumn,letterpaper]{article}

\usepackage[pagenumbers]{wacv} 

\definecolor{wacvblue}{rgb}{0.21,0.49,0.74}
\usepackage[pagebackref,breaklinks,colorlinks,allcolors=wacvblue]{hyperref}
\usepackage{makecell}
\usepackage{float}
\usepackage{placeins}
\def\wacvPaperID{3427} 
\def\confName{WACV}
\def\confYear{2027}

\title{HIVE-Diff: High Frame Rate HDR Video Reconstruction from a Single LDR Frame and Events}

\author{Caixin Wang$^{1,2}$ \quad Qingrui Wang$^{3}$ \quad Jie Zhang$^{2}$ \quad Ralph Etienne-Cummings$^{1}$\\[2pt]
$^{1}$Johns Hopkins University \quad $^{2}$Northwestern University \quad $^{3}$Massachusetts Institute of Technology
}

\begin{document}
\maketitle
\begin{abstract}
High frame rate HDR video is important in applications such as autonomous driving and scientific imaging, where rapid motion and extreme illumination often occur simultaneously. Conventional frame cameras face a trade-off between temporal resolution and dynamic range, while event cameras provide complementary high-speed brightness-change measurements over a wide illumination range. These properties have motivated event-based HDR reconstruction, with existing methods broadly following event-only and hybrid event-frame paradigms. Event-only HDR reconstruction, however, suffers from brightness and color drift due to the lack of absolute intensity references. Hybrid event-frame methods alleviate this ambiguity using low-dynamic-range (LDR) frames as anchors, but most perform one-to-one HDR reconstruction at available LDR timestamps. In this work, we investigate single-reference HDR temporal expansion, reconstructing $N$ HDR frames from one LDR reference and events instead of $N$ aligned LDR observations. We develop HIVE-Diff, an event-conditioned video diffusion framework that decomposes the task into event-guided temporal generation and HDR content completion. Because paired real HDR-event data are scarce, we train on synthetic data and adapt the model to real event domains using LDR-event sequences as proxy supervision with teacher-student distillation. Experiments on synthetic and multiple real-world datasets demonstrate strong reconstruction quality and robust real-data adaptation. Code: \href{https://github.com/caixinw4-afk/hive-diff}{\texttt{github.com/caixinw4-afk/hive-diff}}.
\end{abstract}    
\begin{figure*}[t]
\centering
\includegraphics[height=7.2cm]{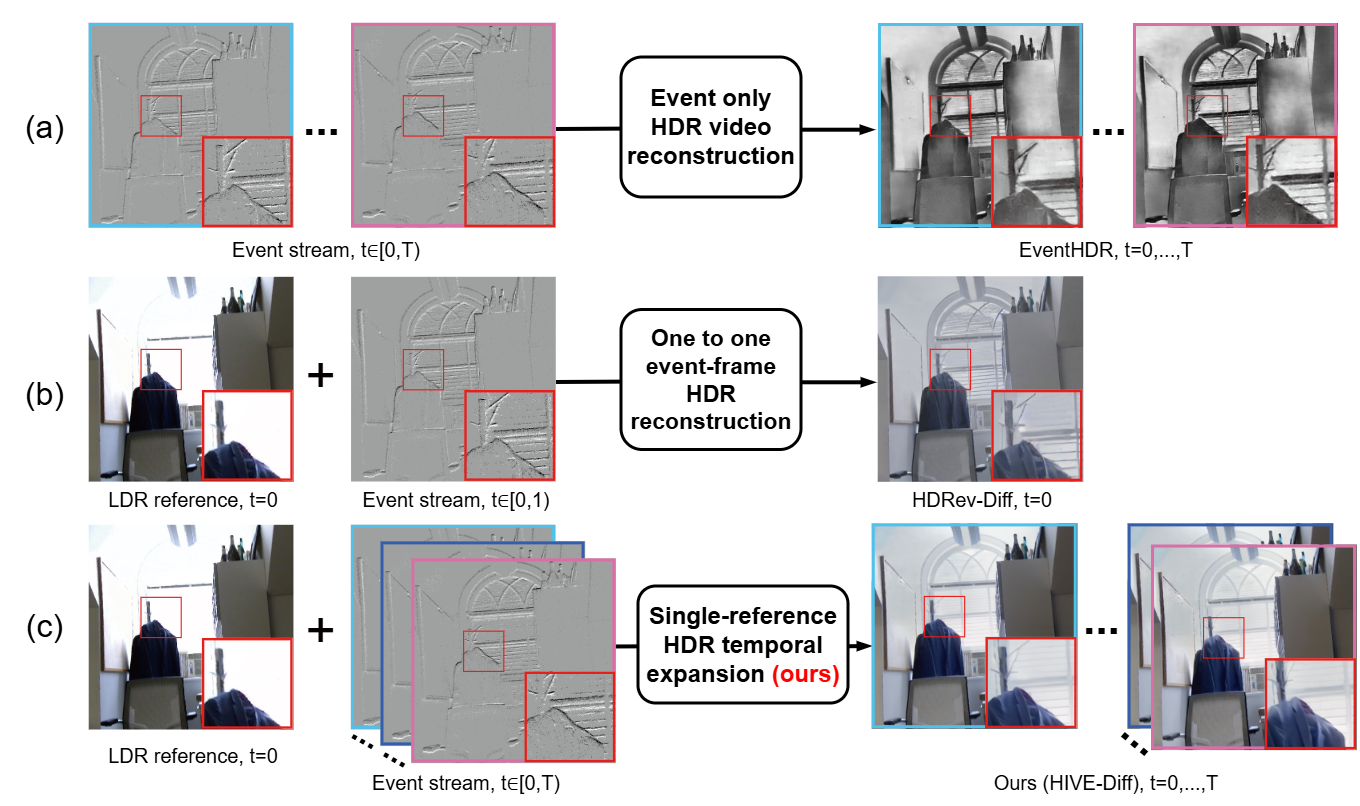}
\caption{Comparison of three event-based HDR reconstruction settings: (a) event-only HDR video reconstruction, represented by EventHDR \cite{zou2024eventhdr}; (b) one-to-one event-frame HDR reconstruction, represented by HDRev-Diff \cite{yang2025event}; and (c) our single-reference HDR temporal expansion, which reconstructs multiple HDR frames from one LDR reference and an event stream.
}
\label{fig:first_figure}
\end{figure*}

\section{Introduction}
\label{sec:intro}

High-dynamic-range (HDR) video is important for capturing scenes with extreme lighting while preserving details in both highlights and shadows. Prior research has explored different methods to capture and reconstruct HDR video, such as multi-exposure fusion, learning-based HDR synthesis, and hardware-assisted sensing \cite{chen2021hdr,kalantari2019deep,cogalan2022learning,wang2024pix2hdr}. Methods using temporally varying exposures with frame alignment are effective for static and slow scenes, but they may fail to reconstruct HDR content at high-speed scenes. Hardware-based approaches avoid inter-frame alignment using spatially varying exposures or multi-sensor designs, but increase system complexity and often sacrifice spatial resolution, field of view, or readout bandwidth \cite{wang2024pix2hdr,nayar2000high,tocci2011versatile}. Together, these limitations make it difficult for conventional frame-based systems to achieve both high dynamic range and high temporal resolution.

Compared with frame-based methods, event cameras record per-pixel brightness changes asynchronously with microsecond latency \cite{gallego2020event,lichtsteiner2008dvs}. This allows them to capture fast motion without blur even at extreme light conditions. Existing event-based HDR methods broadly fall into two categories. Event-only methods reconstruct blur-free, high-speed HDR video directly from event streams. However, this method does not have absolute intensity references and often produces videos with severe color and brightness drifts \cite{wang2019event,zou2021learning,rebecq2019high,zou2024eventhdr}. Hybrid event-frame methods instead use synchronous LDR frames as intensity anchors and uses events to compensate for over-/under-exposed contents. While this approach produces HDR videos with high perceptual quality and color fidelity, most existing methods reconstruct HDR content at the timestamps of the available LDR frames \cite{yang2023learning,han2023hybrid,shaw2022hdr}.

In this work, we explore a different event-frame imaging setting. Instead of reconstructing HDR video either from events alone or through one-to-one HDR reconstruction from individual LDR frames, we ask whether a single LDR frame can serve as the appearance and intensity reference for a temporally expanded HDR sequence, as illustrated in \cref{fig:first_figure}. Specifically, we use the LDR frame to provide absolute color and scene appearance, while the subsequent event stream is used to recover both the temporal evolution of the scene and the missing information in over- or under-exposed regions. In this way, the same event measurements are exploited for both temporal expansion and HDR reconstruction, allowing multiple HDR frames to be generated from only one LDR observation. We refer to this setting as single-reference HDR temporal expansion.

To address this problem, we propose HIVE-Diff, an event-conditioned video diffusion framework for single-reference HDR temporal expansion. We use a pretrained Stable Video Diffusion (SVD) model as the spatiotemporal prior and introduce a ControlNet-style event adaptor \cite{zhang2023adding} to guide the generated dynamics and HDR content. We further decompose training into event-guided LDR video generation and HDR content completion, allowing the model to first learn temporal structure and then recover HDR content.

A practical challenge is adapting the framework to real event-camera data, where paired high-frame-rate HDR-event ground truth is extremely scarce. We therefore train on synthetic event-HDR pairs generated with an event simulator \cite{hu2021v2e}. Simulated events, however, differ from real sensors in noise, contrast thresholds, and event-frame alignment. Since paired LDR videos and event streams are easy to collect, we use them as proxy supervision to fine-tune the event-conditioning modules on real data, with teacher-student distillation preventing feature drift during fine-tuning.

We evaluate our framework on synthetic and multiple real-world event datasets. Since existing event-based HDR methods operate under different input-output settings, we benchmark against representative methods under their respective settings. Experiments show that a single LDR reference with events supports temporally expanded HDR reconstruction, and that the proposed adaptation improves quality and temporal stability across real event datasets. Our main contributions are summarized as follows:

\noindent(1) We investigate single-reference HDR temporal expansion, which jointly reconstructs $N$ HDR frames from one LDR reference and events rather than $N$ aligned LDR observations.

\noindent(2) We develop HIVE-Diff, an event-conditioned video diffusion framework that decomposes this task into event-guided LDR video generation and HDR content completion using a pretrained video prior.

\noindent(3) We introduce an HDR-GT-free real-domain adaptation strategy that uses paired LDR-event data and teacher-student distillation to adapt the model to real event statistics and event-frame misalignment.
\section{Related Work}
\label{sec:related}
\begin{figure*}
\centering
\includegraphics[height=5.9cm]{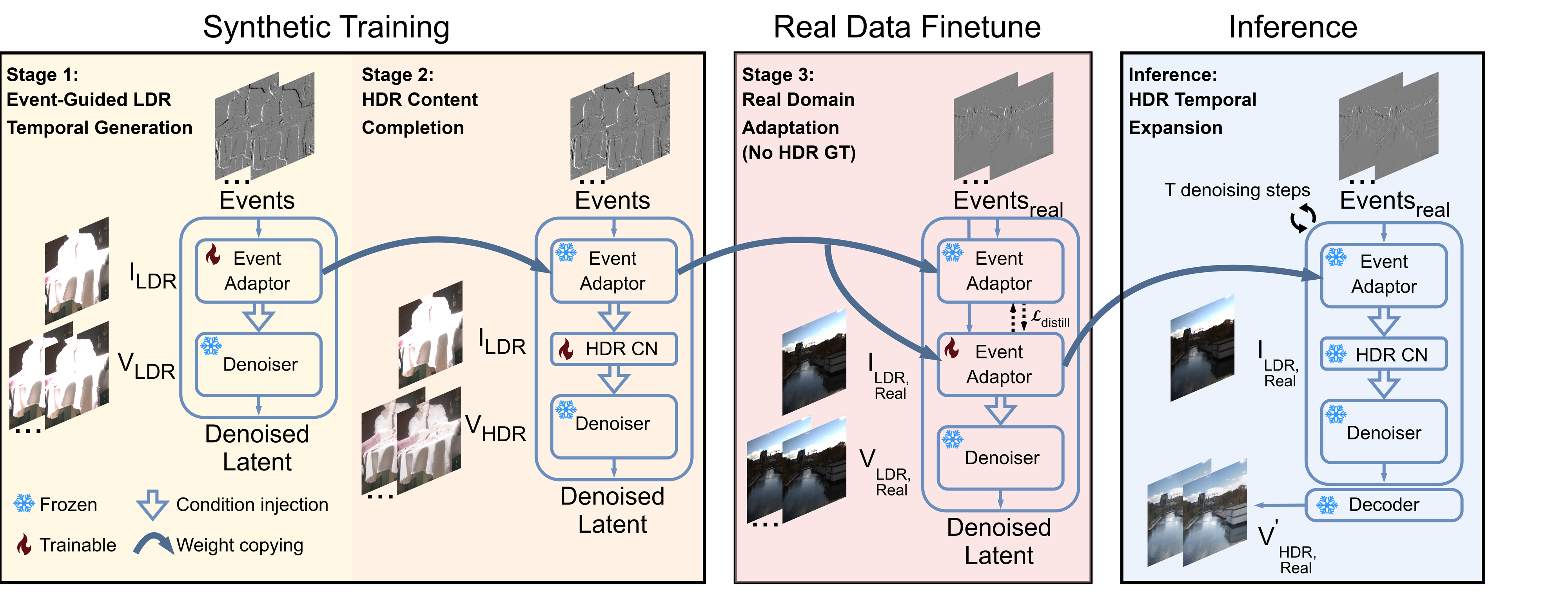}
\caption{
Overview of the HIVE-Diff training and inference pipeline. Stage 1 learns event-guided temporal generation from synthetic LDR-event data, and Stage 2 learns HDR content completion with synthetic HDR supervision. Stage 3 adapts the event adaptor to real event data using LDR-event proxy supervision without HDR ground truth. During inference, the adapted event adaptor is combined with the HDR branch and frozen video diffusion backbone for real-world HDR temporal expansion.
}
\label{fig:pipeline}
\end{figure*}
\subsection{Frame-based HDR Video Reconstruction}

Most frame-based HDR video methods rely on multi-exposure fusion and accurate frame alignment to recover saturated and under-exposed content \cite{kang2003high,zhang2017patch}. However, large motion and unreliable observations in saturated or noisy regions can cause misalignment, ghosting, and temporal inconsistency. Learning-based methods improve alignment through coarse-to-fine estimation and feature-level matching \cite{kalantari2019deep,cogalan2022learning,shu2024towards}, but large motion remains challenging in high-speed scenes. Hardware solutions such as spatially varying or interleaved exposures reduce temporal alignment requirements, but introduce additional sensor complexity and trade-offs in spatial resolution, field of view, or readout bandwidth \cite{cogalan2022learning,wang2024pix2hdr}.
\subsection{Event-based HDR Reconstruction}

Event cameras offer high temporal resolution and a wide effective dynamic range, making them attractive for HDR reconstruction in fast-moving scenes. Event-only methods reconstruct high-speed HDR video directly from events \cite{wang2019event,zou2021learning,rebecq2019high,zou2024eventhdr}, but the lack of absolute intensity observations can cause brightness drift, color ambiguity, and temporal flickering. Hybrid event-frame methods instead use LDR frames as intensity and appearance references and events to recover saturated or under-exposed content \cite{yang2023learning,han2023hybrid,shaw2022hdr,yang2025event,chen2025evhdr}. Most perform one-to-one HDR reconstruction at available LDR timestamps, while several approaches further interpolate between repeated or alternating-exposure observations to produce higher-frame-rate HDR video \cite{chang20231000,samra2023high}. Closer to our setting, AKF~\cite{wang2021akf} interpolates HDR intensity between frames with events, but relies on a continuous frame stream and reduces to drifting event integration when only one frame is available; Zhu~\etal~\cite{zhu2024vfp} and ContinuityCam~\cite{wang2025continuitycam} reconstruct LDR video from one image and the subsequent events, and Lin~\etal~\cite{lin2023rshdr} expand a single rolling-shutter image into HDR frames with events. None of them generates multiple HDR frames from a single LDR frame with a learned video prior.

Unlike these settings, we study single-reference HDR temporal expansion, where one LDR observation and its corresponding event stream are used to reconstruct multiple HDR frames within a temporal window. This setting retains an absolute appearance reference while reducing the need for an aligned LDR observation at every output timestamp.
\subsection{Diffusion Priors for Event-Guided Reconstruction}

Pretrained video diffusion models provide strong spatiotemporal priors for temporally coherent generation from a single image \cite{blattmann2023stable,ho2022video,ho2022imagen}. RE-VDM~\cite{chen2025repurposing} and EGVD~\cite{zhang2025egvd} steer a frozen video diffusion model with events to generate video from a start frame; our Stage~1 builds on this event-conditioning design. Diffusion priors have also been applied to HDR reconstruction and saturated-region recovery \cite{guan2024diffusion,wang2025lediff,yang2025event}, but at existing frame timestamps. HIVE-Diff connects the two with a second ControlNet for HDR content completion and an adaptation stage for real sensors.

\subsection{Real-World Adaptation for Event-based Vision}

Models trained on simulated events often degrade on real sensors because of differences in noise, contrast thresholds, event rates, and other sensor characteristics \cite{gehrig2020video}. Domain and test-time adaptation have therefore been explored for event-based vision \cite{cho2024tta}. For HDR video reconstruction, however, paired real events and high-frame-rate HDR ground truth are rarely available. We therefore adapt the event-conditioning modules with paired real LDR videos and events as proxy supervision (\cref{sec:adapt}).
\section{Proposed Method}
\label{sec:method}
\subsection{Framework Overview}

\begin{figure*}
\centering
\includegraphics[height=6.9cm]{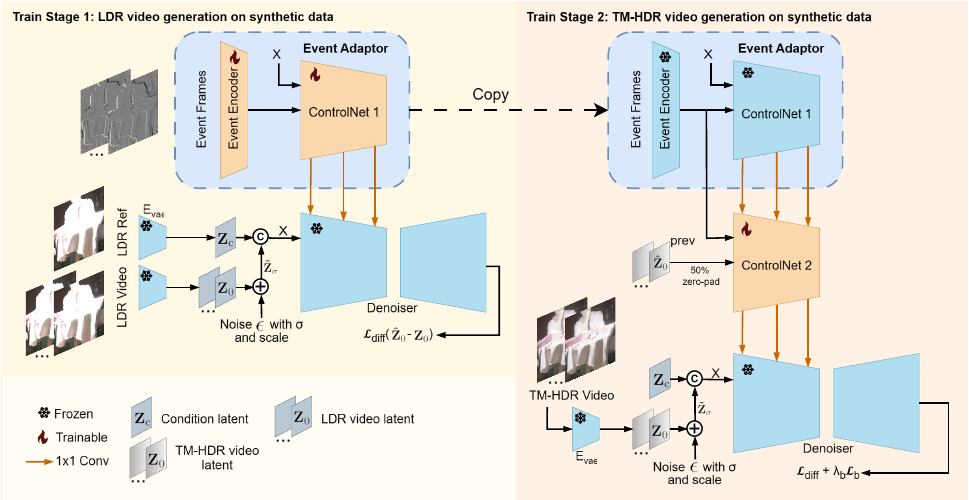}
\caption{Training strategies on synthetic dataset. Stage 1: We encode the randomly exposed LDR reference and the LDR video into latents, add noise to the video latents, and condition a pretrained denoiser via an event adaptor (event encoder + ControlNet 1) to generate LDR video. Stage 2: We freeze a copy of the trained event adaptor, insert the trainable ControlNet 2, and switch the target to normalized PQ-encoded HDR video latents. We also optionally feed previously denoised latents to encourage temporal consistency across video clips}
\label{fig:model}
\end{figure*}

As illustrated in \cref{fig:pipeline}, HIVE-Diff realizes single-reference HDR temporal expansion through three stages: event-guided temporal generation, HDR content completion, and real-domain adaptation. This decomposition learns temporal dynamics and HDR reconstruction from synthetic data, while adapting the event-conditioning module to real sensor characteristics without requiring real HDR ground truth.

In Stage 1, we train an event adaptor on synthetic LDR-event pairs to guide a frozen pretrained video diffusion model in generating an LDR sequence from a single reference. In Stage 2, we freeze the adaptor and train an HDR ControlNet with synthetic HDR supervision to recover HDR content while preserving the learned temporal structure. In Stage 3, we adapt only the event adaptor to real data, using paired real LDR videos and event streams as proxy supervision and a frozen synthetic-trained copy as teacher to prevent feature drift. At inference, the adapted adaptor, the synthetic-trained HDR ControlNet, and the frozen backbone together reconstruct temporally expanded HDR video from real event-camera data.

\subsection{Training on Synthetic Data}

As illustrated in \cref{fig:model}, synthetic training is decomposed into two stages. Stage 1 learns event-guided temporal generation in the LDR domain, while Stage 2 builds on the learned temporal representation for HDR content completion.

\textbf{Stage 1: Event-guided temporal generation.}
We use the pretrained SVD image-to-video denoiser as the video prior and keep its weights frozen to preserve the pretrained spatiotemporal representation. A trainable event adaptor, consisting of an event encoder and a ControlNet-style branch, injects event-derived control signals into the frozen denoiser \cite{chen2025repurposing,yang2025event,zhang2023adding}. For each training clip, we render an LDR video at a randomly sampled exposure and use its first frame as the reference. The LDR video and reference frame are encoded by the pretrained VAE into the target latent $\mathbf{Z}_0$ and reference latent $\mathbf{Z}_c$, respectively. Following the standard SVD/EDM training formulation \cite{karras2022elucidating,blattmann2023stable}, the target latent is perturbed with Gaussian noise, while the reference latent is repeated along the temporal dimension and concatenated with the noisy video latent as input to the denoiser and event adaptor.

The event tensor is built as follows. For each output time $t_k$ we form 4 channels: channel $b\in\{0,\dots,3\}$ is the polarity-signed event count accumulated over a window of length $(2b{+}1)\,\Delta t/4$ centred at $t_k$, where $\Delta t$ is the output frame interval. Stacking these nested windows gives a multi-scale representation \cite{chen2025repurposing,lu2024hr,nam2022stereo}: the shortest window resolves fast motion, the longer ones gather enough events in slowly changing regions. The resulting tensor $\mathbf{V}$ is encoded by a CNN-based event encoder $E(\cdot)$ and fused with the initial ControlNet features:
\begin{align}
    \mathbf{h^{(\mathrm{0})}} \leftarrow \mathbf{h^{(\mathrm{0})}} + E(\mathbf{V}).
\end{align}
The event adaptor then produces multi-scale control residuals through zero-initialized $1\times1$ convolutional layers, which are injected into the down- and mid-blocks of the frozen denoiser. The model is trained using the standard SVD diffusion objective, denoted as $\mathcal{L}_{\mathrm{diff}}$. Through this stage, the event adaptor learns to translate asynchronous event observations into temporal controls for generating an LDR sequence from a single reference frame.

\textbf{Stage 2: HDR content completion.}
After Stage 1, we freeze the trained event adaptor and introduce an HDR ControlNet. For each training clip, an LDR reference is generated from the first HDR frame at a randomly sampled exposure, while the normalized PQ-encoded HDR sequence \cite{froehlich2014creating} serves as the target and is encoded into the clean latent $\mathbf{Z}_0$.

The frozen event features are mapped through zero-initialized $1\times1$ convolutions and injected into the HDR ControlNet, which conditions the same frozen denoiser for HDR content recovery. The exposure value $\alpha$ of the reference is encoded by an MLP and injected through cross-attention; $\alpha$ is the known rendering exposure for synthetic and EventHDR data and is set to $0$ for real captures.

For longer HDR videos, independently synthesized clips may exhibit appearance drift or discontinuities at clip boundaries. To improve long-range temporal consistency, we condition the current clip on the denoised latent from the previous clip and impose a boundary consistency loss between adjacent clips:
\begin{align}
\mathcal{L}_{\mathrm{b}}
&=
\left\|
\hat{\mathbf{Z}}^{\,\mathrm{curr}}_{0,[0]}
-
\hat{\mathbf{Z}}^{\,\mathrm{prev}}_{0,[-1]}
\right\|_2^2,
\\
\mathcal{L}
&=
\mathcal{L}_{\mathrm{diff}}
+
\lambda_{\mathrm{b}}\mathcal{L}_{\mathrm{b}} .
\end{align}
where $[0]$ and $[-1]$ denote the first and last latent frames, respectively.

\subsection{Real-Domain Adaptation without HDR Ground Truth}
\label{sec:adapt}

\begin{figure*}
\centering
\includegraphics[height=6.4cm]{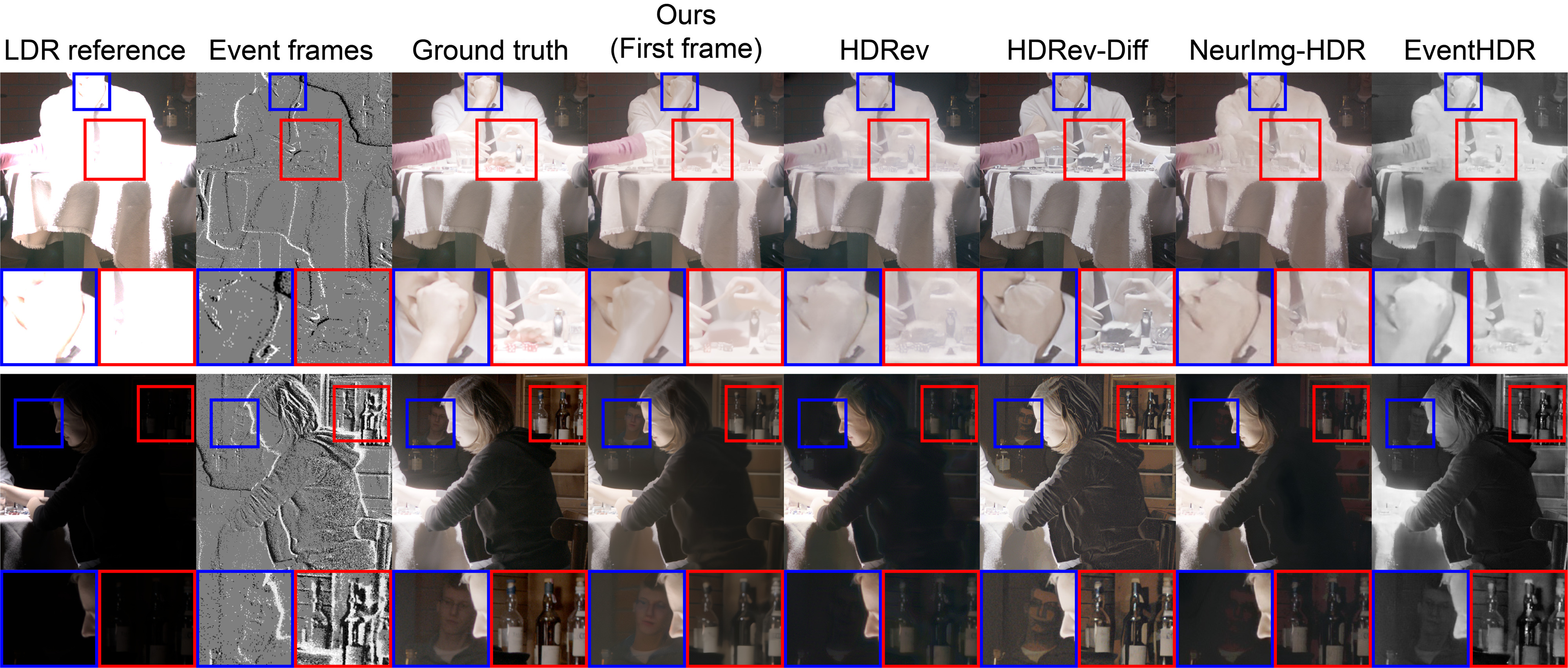}
\caption{Qualitative evaluation of synthetic data on hdm-hdr-2014 \cite{froehlich2014creating} over both over-exposed and under-exposed scenarios}
\label{fig:synthetic_compare}
\end{figure*}

After synthetic training, we adapt only the event adaptor to real data using paired LDR videos and event streams as proxy supervision, while keeping the pretrained denoiser and HDR branch fixed. This allows adaptation to real event statistics and event--frame misalignment without requiring HDR ground truth. To prevent feature drift, we keep a frozen synthetic-trained adaptor as the teacher and fine-tune another copy as the student, as illustrated in Stage 3 of \cref{fig:pipeline}.

In addition to the diffusion loss $\mathcal{L}_{\mathrm{diff}}$, two feature-level distillation losses regularize the student. With $D_{\cos}(\mathbf{a},\mathbf{b})=1-\cos(\mathbf{a},\mathbf{b})$, the event-embedding loss keeps the student embedding $\mathbf{E}_{\mathrm{s}}$ close to the teacher embedding $\mathbf{E}_{\mathrm{t}}$, and the residual loss does the same for the multi-scale control residuals $\mathbf{R}^{(d)}$ at the selected down- and mid-blocks $d\in D$:
\begin{align}
    \mathcal{L}_{\mathrm{e}} &= \tfrac{1}{|K|}\textstyle\sum_{k \in K} D_{\cos}\big(\mathbf{E}_{\mathrm{s}}^{(k)},\mathbf{E}_{\mathrm{t}}^{(k)}\big),\\
    \mathcal{L}_{\mathrm{r}} &= \tfrac{1}{|D|}\textstyle\sum_{d \in D} D_{\cos}\big(\mathbf{R}_{\mathrm{s}}^{(d)},\mathbf{R}_{\mathrm{t}}^{(d)}\big),
\end{align}
where $K$ indexes the flattened event features. The final adaptation objective is $\mathcal{L}_{\mathrm{f}}=\mathcal{L}_{\mathrm{diff}}+\lambda_{\mathrm{e}}\mathcal{L}_{\mathrm{e}}+\lambda_{\mathrm{r}}\mathcal{L}_{\mathrm{r}}$.

\subsection{Training Data Synthesis and Implementation Details}

We use the 16-bit color-graded HDR TIFF sequences from HdM-HDR-2014 \cite{froehlich2014creating} as the source HDR data. The PQ-encoded pixel values are normalized to $[0,1]$ and used as the HDR supervision. Following the exposure-rendering procedure provided with the dataset, we synthesize LDR observations at randomly sampled exposure values. We further apply randomly sampled global camera motion to the video sequences so that static scene regions also produce temporal intensity changes, and generate the corresponding event streams using the v2e event simulator \cite{hu2021v2e}.

We train two variants of HIVE-Diff with clip lengths of 5 and 20 frames at $512\times512$ resolution with 4 event bins on a single NVIDIA L40S GPU. Optimizer settings, learning rates, and step counts are given in the supplementary material.

\section{Experiments}
\label{sec:Experiments}

Because no existing method jointly addresses single-reference HDR temporal expansion, we compare representative event-based HDR methods. EventHDR~\cite{zou2024eventhdr} uses events only, whereas NeurImg-HDR~\cite{han2023hybrid}, HDRev~\cite{yang2023learning}, and HDRev-Diff~\cite{yang2025event} perform one-to-one LDR-event HDR reconstruction. An input-matched cascade would first predict the missing LDR frames and make its performance depend on both LDR generation and HDR reconstruction. We therefore evaluate each one-to-one method under its native and more favorable protocol—an aligned LDR input at every output timestamp—while HIVE-Diff uses only the initial frame. We also tried an input-matched variant of EventHDR that fits a per-window affine brightness and color correction of its output to the LDR reference at $t{=}0$. On EventHDR this raises its PSNR by only about 1\,dB and changes LPIPS little, because the anchor corrects the global level but not the structural errors of the event-only reconstruction; we therefore do not report it as a separate row. We report F$^\ast$ for the reference-aligned first output frame, and 5$\times$/20$\times$ for full-sequence averages of the variants that generate 5 or 20 HDR frames per LDR reference. All baselines are retrained on the same synthetic data with identical LDR rendering and event simulation protocols using official implementations and recommended settings.

For datasets with HDR ground truth, we report PSNR \cite{gonzalez2009digital},  Scale- and Shift-Invariant PSNR (SSI-PSNR)~\cite{weigert2018content,krull2020probabilistic}, SSIM~\cite{wang2004image}, LPIPS~\cite{zhang2018unreasonable}, CIEDE2000~\cite{luo2001development}, and temporal warping error ($E_{\mathrm{warp}}$)~\cite{lai2018learning}. $E_{\mathrm{warp}}$ is undefined for F$^\ast$. HdM-HDR-2014 metrics are computed against normalized Perceptual Quantizer (PQ)-encoded HDR ground truth. For datasets without HDR ground truth, we report MANIQA \cite{yang2022maniqa} on F$^\ast$, and Event-Photometric Consistency (EPC) on 5$\times$ outputs. EPC checks whether the reconstructed brightness changes agree with the recorded events. With $\hat{L}_t$ the log-luminance computed from the PQ-encoded output $\hat{I}_t$ and $E_t$ the net polarity count of the real events between $t$ and $t{+}1$, the ideal event model gives $\hat{L}_{t+1}-\hat{L}_t\approx C\,E_t$ for contrast threshold $C$ \cite{gallego2020event,rebecq2018esim}; we therefore report
\begin{equation}
  \mathrm{EPC}=\tfrac{1}{|\mathcal{T}|}\textstyle\sum_{t\in\mathcal{T}}\rho\!\left(G_\sigma * (\hat{L}_{t+1}-\hat{L}_t),\; G_\sigma * E_t\right),
\end{equation}
with $\rho$ the Pearson correlation and $G_\sigma$ a Gaussian kernel ($\sigma{=}2$\,px) that absorbs small misalignment and sensor noise. EPC does not require HDR ground truth. It measures how well brightness changes match the events, regardless of the overall brightness scale or offset. Additional multi-timestamp results and frame-wise quality curves are provided in the supplementary material.

\begin{figure*}[t]
\centering

\begin{subfigure}[t]{0.98\textwidth}
    \centering
    \includegraphics[height=3.9cm,keepaspectratio]{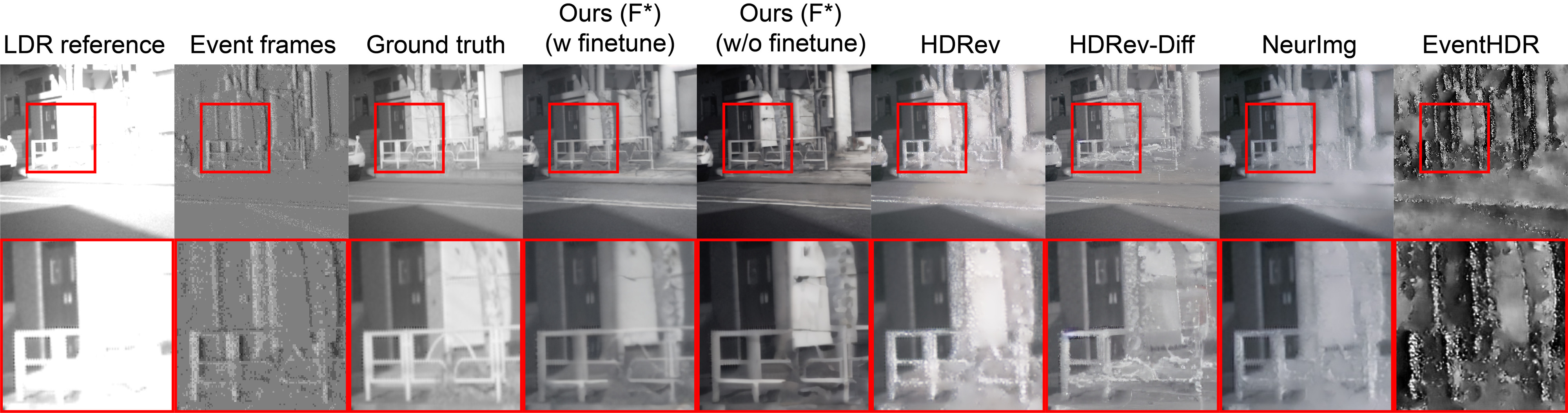}
    \caption{EventHDR dataset \cite{zou2024eventhdr}.}
    \label{fig:eventhdr_compare}
\end{subfigure}

\begin{subfigure}[t]{0.98\textwidth}
    \centering
    \includegraphics[height=9.0cm,keepaspectratio]{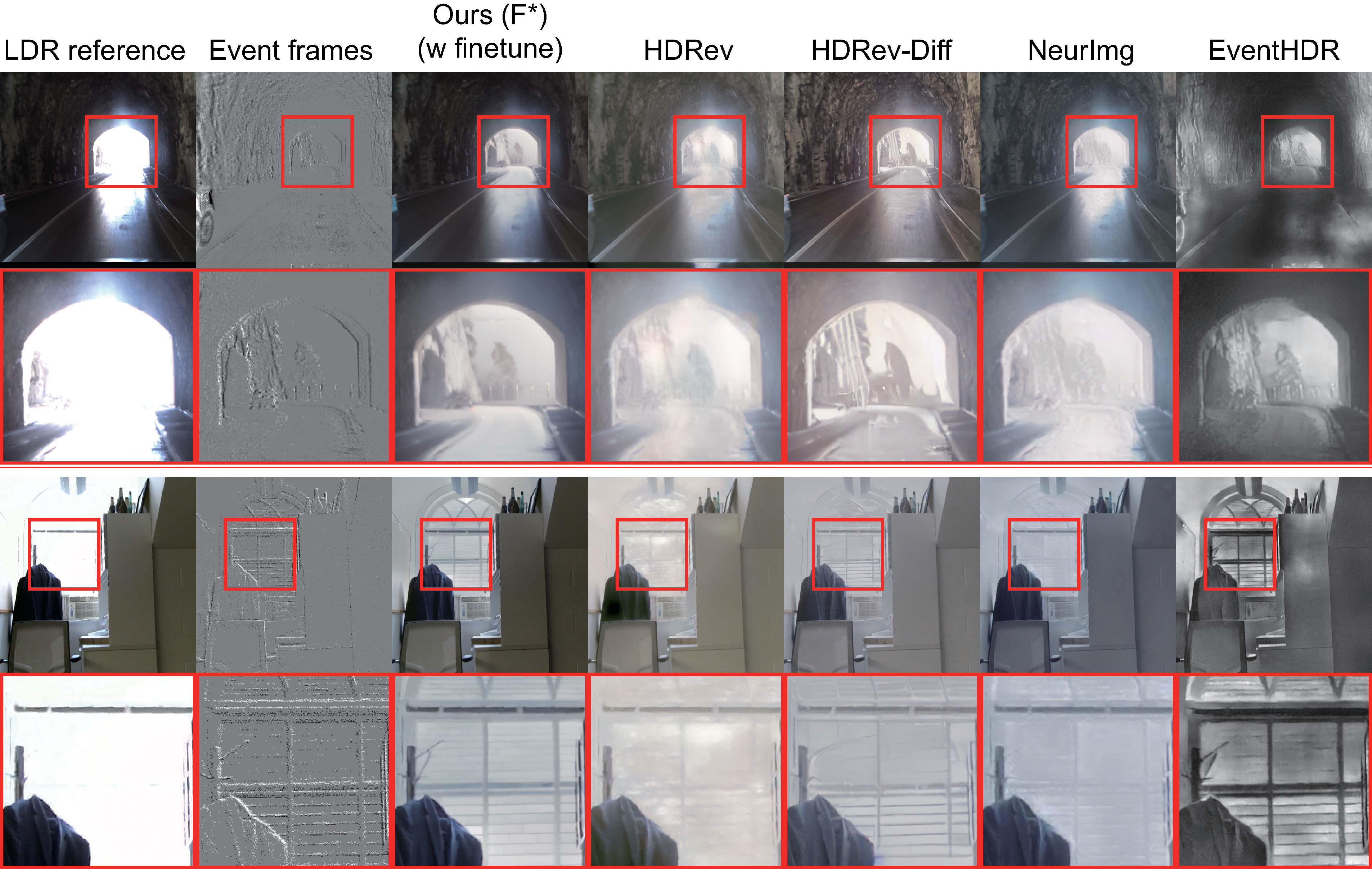}
    \caption{DSEC \cite{Gehrig21ral} (top) and our self-collected data (bottom); HS-ERGB \cite{Tulyakov21CVPR} results are in the supplementary material.}
    \label{fig:real_compare}
\end{subfigure}

\vspace{-1mm}

\caption{
Qualitative evaluation on real event datasets.
(a) EventHDR provides HDR ground truth for evaluating real-domain adaptation.
(b) DSEC and our self-collected data evaluate cross-sensor real-domain adaptation without HDR ground truth.
}
\label{fig:real_evaluation}
\end{figure*}

\subsection{Controlled Evaluation on Synthetic Data}

\begin{table}
  \centering
  \setlength{\tabcolsep}{1.3pt}
  \resizebox{\columnwidth}{!}{%
  \begin{tabular}{@{}lccccccc@{}}
    \toprule
    Methods & \#Ref & PSNR$\uparrow$ & SSI-PSNR$\uparrow$ & SSIM$\uparrow$ & LPIPS$\downarrow$ & CIEDE$\downarrow$ & $E_{\mathrm{warp}}\downarrow$\\
    \midrule

    EventHDR \cite{zou2024eventhdr}
    & 0 & 19.60 & 21.28 & 0.774 & 0.257 & 9.20 & 0.0189\\

    \midrule

    NeurImg-HDR \cite{han2023hybrid}
    & $N$ & 20.42 & 25.32 & \textcolor{blue}{0.806} & 0.243 & 7.76 & 0.0099\\

    HDRev \cite{yang2023learning}
    & $N$ & 20.73 & 25.13 & 0.763 & 0.267 & 8.99 & 0.0092\\

    HDRev-Diff \cite{yang2025event}
    & $N$ & 18.97 & 25.67 & 0.692 & \textcolor{blue}{0.229} & 9.76 & 0.0365\\

    Ours (F$^\ast$)
    & 1 & \textcolor{red}{21.85}
    & \textcolor{red}{25.99}
    & \textcolor{red}{0.814}
    & \textcolor{red}{0.224}
    & \textcolor{red}{6.59}
    & -- \\

    \midrule

    Ours (5x)
    & 1 & \textcolor{blue}{21.71} & \textcolor{blue}{25.71} & 0.802 & 0.237 & \textcolor{blue}{6.65} & \textcolor{red}{0.0059}\\

    Ours (20x)
    & 1 & 18.83 & 25.48 & 0.735 & 0.238 & 8.65 & \textcolor{blue}{0.0091}\\

    \bottomrule
  \end{tabular}
  }
    \caption{Controlled evaluation on the held-out ``poker\_fullshot'' sequence of HdM-HDR-2014 \cite{froehlich2014creating}, averaged over over- and under-exposed scenarios. \#Ref: LDR reference frames needed for an $N$-frame output. F$^\ast$: reference-aligned first frame; 5$\times$/20$\times$: full-sequence averages. $E_{\mathrm{warp}}$ is undefined for F$^\ast$. Red/blue: best/second best.}
  \label{tab:synthetic_comparison}
\end{table}

\begin{table}
  \centering
  \setlength{\tabcolsep}{1.3pt}
  \resizebox{\columnwidth}{!}{%
  \begin{tabular}{@{}lccccccc@{}}
    \toprule
    Methods & \#Ref & PSNR$\uparrow$ & SSI-PSNR$\uparrow$ & SSIM$\uparrow$ & LPIPS$\downarrow$ & CIEDE$\downarrow$ & $E_{\mathrm{warp}}\downarrow$\\
    \midrule

    EventHDR \cite{zou2024eventhdr}
    & 0 & 11.22 & 12.34 & 0.353 & 0.690 & 20.63 & 0.0885 \\

    \midrule

    NeurImg-HDR \cite{han2023hybrid}
    & $N$ & 17.69 & 23.15 & 0.803 & 0.259 & 12.07 & \textcolor{blue}{0.0141} \\

    HDRev \cite{yang2023learning}
    & $N$ & 17.09 & 21.20 & 0.795 & 0.349 & \textcolor{blue}{12.05} & 0.0148 \\

    HDRev-Diff \cite{yang2025event}
    & $N$ & 15.19 & 17.86 & 0.518 & 0.566 & 14.18 & 0.0568 \\

    Ours (w/o FT, F$^\ast$)
    & 1 & 16.10 & 23.11 & 0.811 & 0.212 & 13.77 & -- \\

    Ours (w/ FT, F$^\ast$)
    & 1 & \textcolor{red}{18.03} & \textcolor{red}{23.99} & \textcolor{red}{0.830} & \textcolor{red}{0.190} & 12.07 & -- \\

    \midrule

    Ours (w/ FT, 5x)
    & 1 & \textcolor{blue}{17.98} & \textcolor{blue}{23.93} & \textcolor{blue}{0.825} & \textcolor{blue}{0.197} & 12.08 & \textcolor{red}{0.0065} \\

    Ours (w/ FT, 20x)
    & 1 & 17.37 & 22.58 & 0.819 & 0.224 & \textcolor{red}{11.30} & \textcolor{red}{0.0065} \\

    \bottomrule
  \end{tabular}
  }
    \caption{ Results on five held-out EventHDR sequences \cite{zou2024eventhdr}. FT uses 19 disjoint LDR-event sequences without HDR GT; applying this objective to the baselines would require modifying their original formulations. }
  \label{tab:eventhdr_compare}
\end{table}

We first conduct a controlled synthetic evaluation on the held-out ``poker\_fullshot'' sequence from HdM-HDR-2014 \cite{froehlich2014creating}, where HDR ground truth enables direct assessment of reconstruction fidelity and temporal consistency; broader real-world generalization is evaluated in \cref{sec:real_world_evaluation}. As shown in \cref{tab:synthetic_comparison}, under the first-frame protocol (F$^\ast$), our method achieves the best performance across all frame-wise metrics compared with the one-to-one hybrid HDR baselines. The 5$\times$ model remains close to F$^\ast$, while 20$\times$ shows larger degradation in PSNR, SSIM and CIEDE2000. For temporal consistency, the 5$\times$ and 20$\times$ variants achieve $E_{\mathrm{warp}}$ values of 0.0059 and 0.0091, respectively, ranking first and second among the evaluated methods.

\cref{fig:synthetic_compare} further shows qualitative comparisons under over- and under-exposed conditions. EventHDR \cite{zou2024eventhdr} produces grayscale reconstructions without reliable color information. HDRev \cite{yang2023learning} and NeurImg-HDR \cite{han2023hybrid} recover coarse structures when events are sufficiently informative, but exhibit color or brightness distortions in severely exposed regions and degrade when event information is sparse. HDRev-Diff \cite{yang2025event} generates perceptually plausible content, although some reconstructed details deviate from the ground truth. In comparison, our method better preserves color and structural fidelity while reconstructing multiple HDR frames from the same LDR reference and event stream.

\subsection{Real-World Evaluation with HDR Ground Truth}
\label{sec:real_world_evaluation}

We first evaluate real-world performance on the EventHDR dataset \cite{zou2024eventhdr}, which contains paired high-frame-rate grayscale HDR videos and real event streams. We render over-exposed LDR clips from the HDR frames at EV=+4 and use the first LDR frame of each temporal window as the reference. To evaluate generalization after real-domain adaptation, we use 19 sequences exclusively for fine-tuning and five disjoint, unseen sequences for evaluation. Fine-tuning uses the rendered LDR sequences and corresponding real events, while the original HDR values are not used as reconstruction targets. Compared with the synthetic data, EventHDR exhibits different event statistics and imperfect event--frame alignment, allowing us to evaluate adaptation to real sensor characteristics.

Quantitative results are reported in \cref{tab:eventhdr_compare}. Fine-tuning with LDR-event proxy supervision consistently improves reconstruction quality on the five unseen test sequences. For Ours (F$^\ast$), fine-tuning improves all frame-wise metrics, including PSNR, SSI-PSNR, SSIM, LPIPS, and CIEDE2000. The adapted 5$\times$ model maintains frame-wise quality close to F$^\ast$ while achieving the best temporal consistency. The 20$\times$ model supports a longer temporal horizon with some degradation in frame-wise quality, while retaining the second-best temporal consistency. Both variants outperform the baselines in temporal consistency despite using only a single LDR reference per temporal window, demonstrating generalization to unseen real sequences.

\cref{fig:eventhdr_compare} shows qualitative results on held-out EventHDR sequences. The synthetic-only model exhibits brightness deviations, blur, and ghosting due to domain shift. After adaptation, these artifacts are substantially reduced, producing sharper reconstructions more consistent with the HDR ground truth and improved temporal consistency.

\subsection{Evaluation on Real Data without HDR Ground Truth}

We further evaluate our method on DSEC \cite{Gehrig21ral}, HS-ERGB \cite{Tulyakov21CVPR}, and our self-collected event-frame dataset, which provide paired LDR RGB videos and events but no HDR ground truth. For each dataset, we use non-overlapping adaptation and evaluation segments. The event adaptor is fine-tuned on paired LDR-event data from the adaptation segments, while all reported results are computed on held-out segments unseen during fine-tuning. This enables real-domain adaptation without HDR supervision.

Qualitative results are shown in \cref{fig:real_compare}. On DSEC, our method better recovers structures in the over-exposed sky when the vehicle exits the tunnel. Similar improvements are observed on our self-collected data and on HS-ERGB (supplementary material), where our method better preserves structure and color in severely over-exposed regions. Without HDR ground truth, \cref{table:real_compare} reports MANIQA on F$^\ast$ for perceptual quality and EPC over five-frame outputs for event consistency. A higher EPC indicates that reconstructed temporal changes better follow the sensor-recorded scene dynamics rather than model hallucinations. Fine-tuning improves EPC from 0.716 to 0.730, while MANIQA decreases slightly from 0.272 to 0.261; nevertheless, our two variants rank first and second on both metrics. Generated videos and additional results are provided in the supplementary material.

\begin{table}[tb]
  \centering
    \caption{
    Evaluation without HDR ground truth. MANIQA is evaluated on F$^\ast$, while EPC is evaluated over five-frame sequences.
    }
  \label{table:real_compare}

  \small
  \setlength{\tabcolsep}{5pt}
  \begin{tabular}{@{}lccc@{}}
    \toprule
    Method & \#Ref & MANIQA$\uparrow$ & EPC$\uparrow$ \\
    \midrule

    EventHDR \cite{zou2024eventhdr}
    & 0 & 0.240 & 0.535 \\

    NeurImg-HDR \cite{han2023hybrid}
    & $N$ & 0.220 & 0.660 \\

    HDRev \cite{yang2023learning}
    & $N$ & 0.228 & 0.610 \\

    HDRev-Diff \cite{yang2025event}
    & $N$ & 0.253 & 0.627 \\

    Ours (w/o FT)
    & 1 & \textcolor{red}{0.272} & \textcolor{blue}{0.716} \\

    Ours (w/ FT)
    & 1 & \textcolor{blue}{0.261} & \textcolor{red}{0.730} \\

    \bottomrule
  \end{tabular}
\end{table}

\section{Ablation Study}

We evaluate the effects of real-domain fine-tuning and teacher--student distillation on the EventHDR dataset \cite{zou2024eventhdr}, as summarized in \cref{table:ablation_compare}. Without fine-tuning, the model is less adapted to real event statistics and event--frame misalignment. Fine-tuning without distillation improves SSI-PSNR and temporal consistency, but degrades PSNR, SSIM, LPIPS, and CIEDE2000, consistent with feature drift during adaptation. Adding teacher--student distillation substantially improves PSNR and overall reconstruction fidelity while preserving the same temporal warping error. Although the model without distillation obtains a higher SSI-PSNR, the full model achieves the best PSNR and better overall quality. Additional qualitative analyses of real-domain fine-tuning and ablations on previous-latent conditioning are provided in the supplementary material.

\begin{table}[tb]
  \centering
  \caption{
  Ablation study of real-domain adaptation on EventHDR
  \cite{zou2024eventhdr}. Frame-wise metrics are reported at
  F$^\ast$, while $E_{\mathrm{warp}}$ is computed over the
  corresponding 5-frame sequence.
  }
  \label{table:ablation_compare}
  \setlength{\tabcolsep}{2pt}

  \resizebox{\columnwidth}{!}{%
  \begin{tabular}{@{}lcccccc@{}}
    \toprule
    Method
    & PSNR$\uparrow$
    & SSI-PSNR$\uparrow$
    & SSIM$\uparrow$
    & LPIPS$\downarrow$
    & CIEDE$\downarrow$
    & $E_{\mathrm{warp}}\downarrow$ \\
    \midrule

    Ours (w/ FT)
    & \textcolor{red}{18.03}
    & \textcolor{blue}{23.99}
    & \textcolor{red}{0.830}
    & \textcolor{red}{0.190}
    & \textcolor{red}{12.07}
    & \textcolor{red}{0.0065} \\

    Ours (w/o distill.)
    & 15.54
    & \textcolor{red}{24.71}
    & 0.805
    & 0.238
    & 16.65
    & \textcolor{red}{0.0065} \\

    Ours (w/o FT)
    & \textcolor{blue}{16.10}
    & 23.11
    & \textcolor{blue}{0.811}
    & \textcolor{blue}{0.212}
    & \textcolor{blue}{13.77}
    & \textcolor{blue}{0.0070} \\

    \bottomrule
  \end{tabular}%
  }
\end{table}

\section{Limitations and Conclusion}

\begin{table}[t]
  \centering
  \caption{Inference cost at $512\times512$ on one NVIDIA L40S GPU, per output frame. Ours: 5$\times$ model, 25 Euler steps, FP16, CFG. HDRev-Diff: 50 DDIM steps, FP32, CFG. Other baselines: FP32 feed-forward.}
  \label{tab:runtime}
  \small
  \setlength{\tabcolsep}{2pt}
  \begin{tabular}{@{}lcccc@{}}
    \toprule
    Method & Params (M) & Time (s) & TFLOPs & Mem.\ (GiB) \\
    \midrule
    HDRev \cite{yang2023learning}        & 57.9 & 0.037 & 1.6 & 1.4 \\
    NeurImg-HDR \cite{han2023hybrid}     & 44.0 & 0.052 & 1.1 & 1.1 \\
    EventHDR \cite{zou2024eventhdr}      & 3.3  & 0.152 & 2.6 & 2.0 \\
    HDRev-Diff \cite{yang2025event}      & 1494 & 4.15  & 120.4 & 6.7 \\
    Ours                                 & 3620 & 1.79  & 120.6 & 9.2 \\
    \bottomrule
  \end{tabular}
\end{table}

HIVE-Diff relies on iterative diffusion denoising, and \cref{tab:runtime} reports its inference cost. Our method is about 2.3$\times$ faster per frame than HDRev-Diff, but still around 100$\times$ slower than feed-forward methods. Few-step distillation and feature caching are promising ways to close this gap.

Beyond inference cost, HIVE-Diff has four limitations. First, fidelity decreases gradually with distance from the reference: the 20$\times$ model loses about 2\,dB in SSI-PSNR from the first to the last output frame on both HdM-HDR-2014 and EventHDR (per-frame curves in the supplementary material). Second, where a saturated region produces no events, such as a static blown-out sky, its content comes from the video prior alone and is plausible rather than measured. Third, previous-latent conditioning smooths clip boundaries, but brightness and color drift still accumulate over long sequences. Fourth, adaptation is performed once per sensor with a few LDR-event sequences; generalization to an unseen camera without adaptation is not evaluated.

In this work, we studied single-reference HDR temporal expansion, reconstructing multiple HDR frames from one LDR reference and events, and developed HIVE-Diff, an event-conditioned video diffusion framework that combines event-guided temporal generation with HDR content completion and adapts to real sensors through proxy-supervised, distillation-regularized fine-tuning without HDR ground truth. Experiments on synthetic and real event datasets show strong reconstruction quality, temporal consistency, and robust real-domain adaptation.

{
    \small
    \bibliographystyle{ieeenat_fullname}
    \bibliography{main}
}

\clearpage
\onecolumn
\setcounter{section}{0}
\renewcommand{\thesection}{\Alph{section}}
\begin{center}
{\Large\bf Supplementary Material}\\[14pt]
\end{center}
\section{Overview}

This appendix provides additional analyses and qualitative
results for HIVE-Diff, including implementation details, real-domain
adaptation, qualitative results on HS-ERGB, frame-wise temporal
evaluation, previous-latent conditioning for cross-clip continuity,
additional qualitative results, and a discussion of inference cost.

\section{Implementation Details}
\label{sec:supp_impl}

\textbf{Training.} All training stages use AdamW~\cite{loshchilov2017decoupled}
with $\beta_1=0.9$, $\beta_2=0.999$, a weight decay of $10^{-2}$, a batch
size of 1, and gradient accumulation over 16 steps. Stages~1 and~2 use a
learning rate of $1\times10^{-4}$ and are trained for 9,000 and 16,000
steps, respectively. Real-domain adaptation uses a learning rate of
$1\times10^{-5}$ for 1,000 steps. The 5$\times$ and 20$\times$ variants
share these settings and are trained at $512\times512$ resolution with
4 event bins on a single NVIDIA L40S GPU.

\textbf{Event-Photometric Consistency.} In the EPC definition of the
main paper, the log-luminance of an output frame is
$\hat{L}_t=\log(Y_t^{\gamma}+\epsilon)$, where $Y_t\in[0,1]$ is the luminance
of the PQ-encoded output $\hat{I}_t$, the power $\gamma=2.2$ expands the PQ
code values toward a linear scale before the log, and $\epsilon=10^{-3}$
avoids $\log 0$. The same transform is applied to all methods. $E_t$ is the per-pixel
signed count of the real events recorded between the timestamps of
outputs $t$ and $t{+}1$. Both maps are smoothed with a Gaussian kernel of
$\sigma=2$\,px, the Pearson correlation is computed over the frame, and
the result is averaged over all consecutive frame pairs of the 5$\times$
outputs.

\section{Effect of Real-Domain Adaptation}
\label{sec:supp_real_adaptation}

\begin{figure}[H]
    \centering
    \includegraphics[width=0.98\textwidth]
    {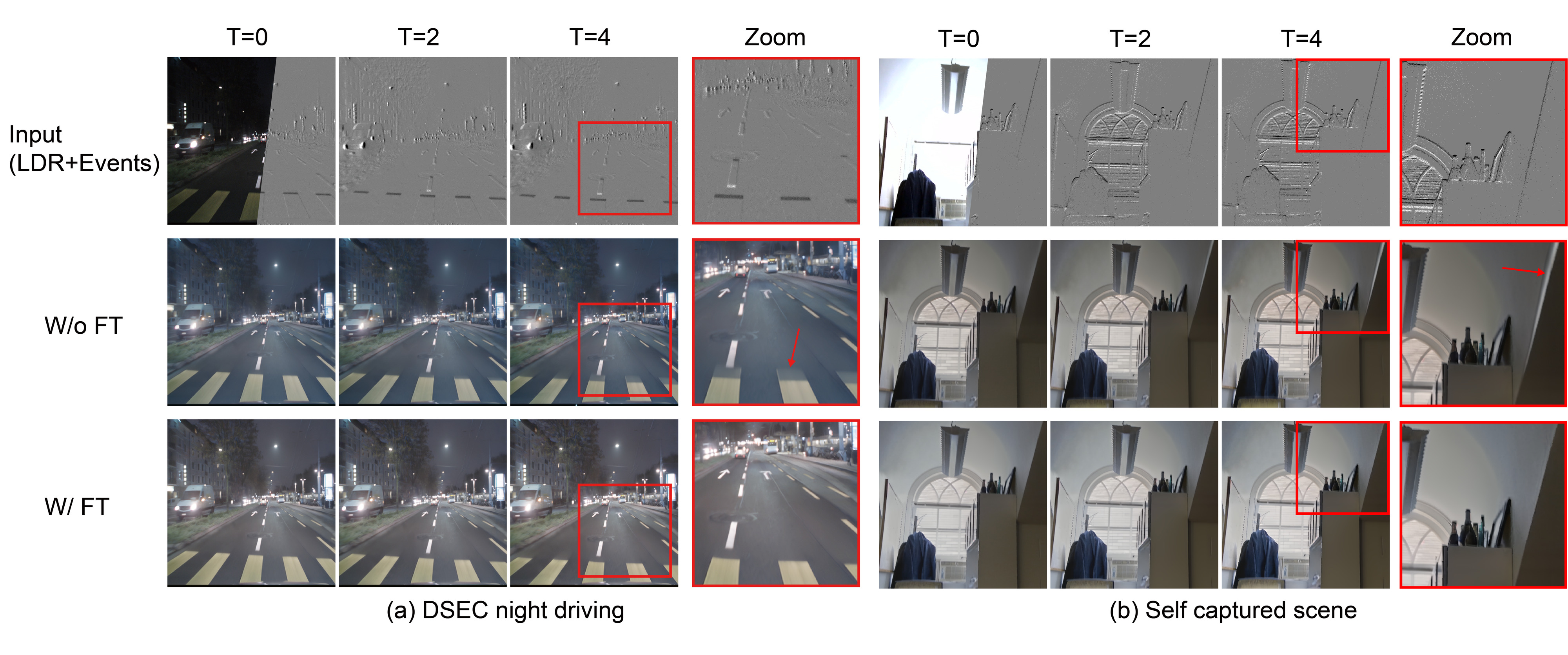}
    \caption{
    Effect of real-domain adaptation on (a) DSEC~\cite{Gehrig21ral}
    and (b) our self-collected data. The top row shows the LDR/event
    inputs, while the middle and bottom rows show outputs without and
    with real-domain fine-tuning, respectively. Red arrows highlight
    representative ghosting artifacts caused by the sim-to-real gap
    when the synthetic-trained model is directly applied to real data.
    Fine-tuning with real LDR-event pairs as proxy supervision
    substantially suppresses these artifacts.
    }
    \label{fig:real_finetune_compare}
\end{figure}

We visualize the effect of real-domain adaptation in
\cref{fig:real_finetune_compare}. When the synthetic-trained model is
directly applied to real event data, the sim-to-real gap leads to
noticeable ghosting artifacts, as highlighted by the red arrows in the
figure. Fine-tuning the event adaptor with paired real LDR-event data as
proxy supervision substantially suppresses these artifacts and improves
structural consistency across frames, as observed on both
DSEC~\cite{Gehrig21ral} and our self-collected data.

\FloatBarrier

\section{Qualitative Results on HS-ERGB}
\label{sec:supp_hsergb}

\begin{figure}[H]
    \centering
    \includegraphics[width=\textwidth]{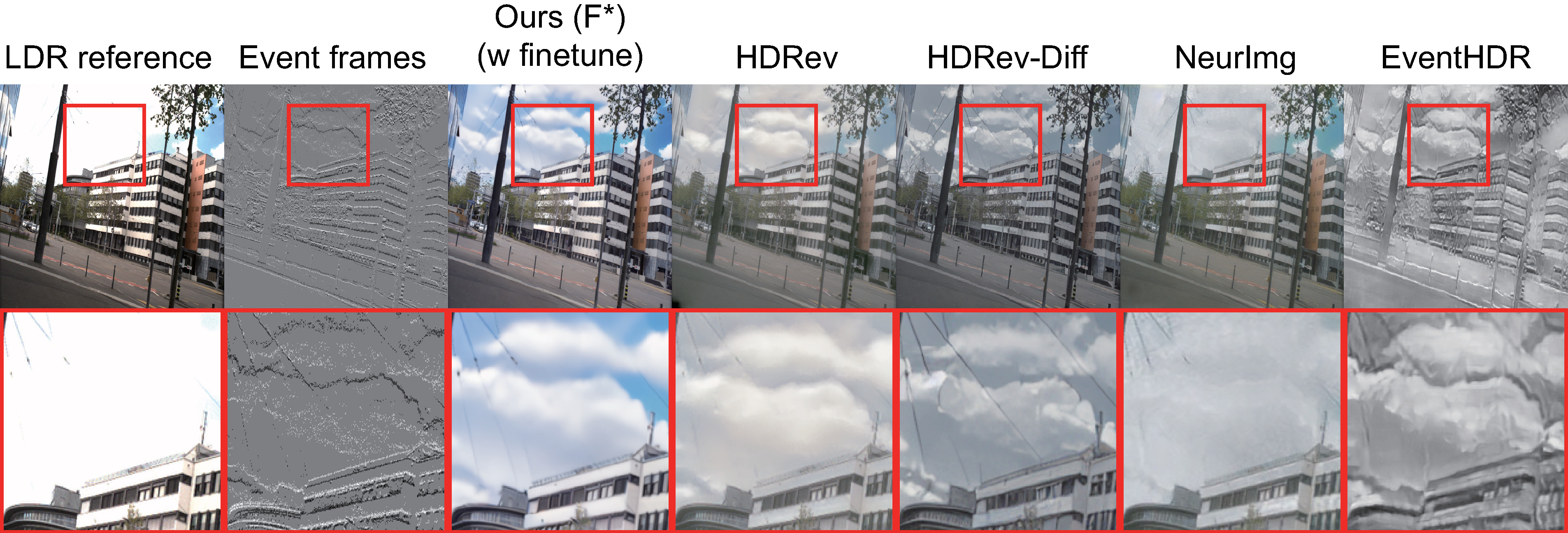}
    \caption{Qualitative comparison on HS-ERGB~\cite{Tulyakov21CVPR}
    (reference-aligned output F$^\ast$). Columns follow Fig.~5 of the main
    paper.}
    \label{fig:supp_hsergb}
\end{figure}

\cref{fig:supp_hsergb} shows the HS-ERGB~\cite{Tulyakov21CVPR} comparison
that accompanies Fig.~5(b) of the main paper. As on DSEC and our
self-collected data, HIVE-Diff recovers structure in the over-exposed
regions while preserving the color of the LDR reference.

\FloatBarrier

\section{Temporal Evaluation over the Reconstruction Horizon}
\label{sec:supp_temporal_evaluation}

\begin{figure}[H]
    \centering
    \includegraphics[width=\textwidth]{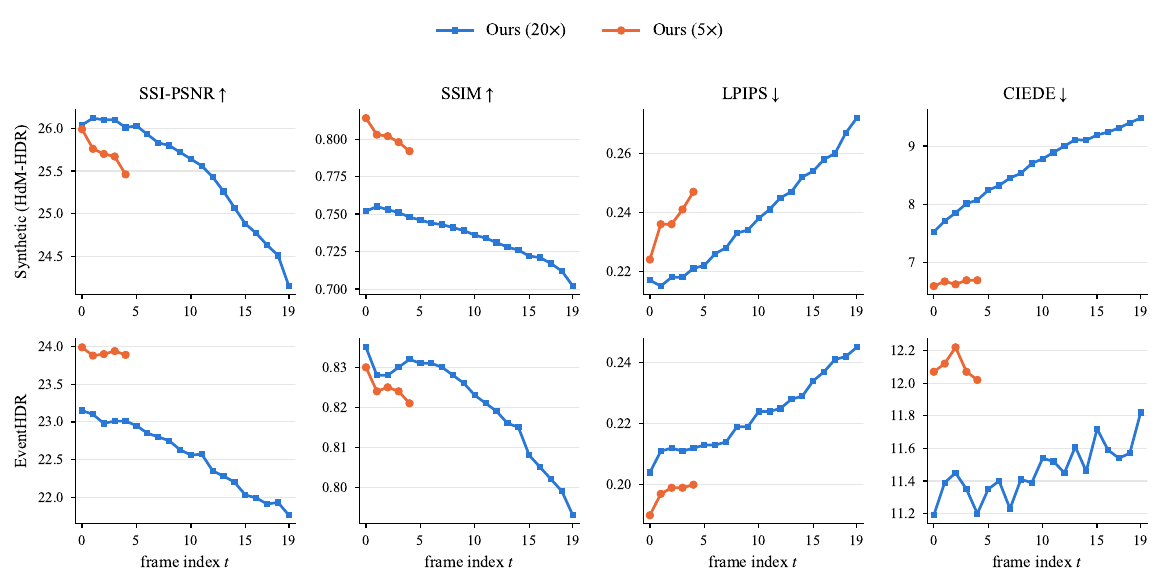}
    \caption{
    Frame-wise reconstruction quality over the temporal horizon on
    HdM-HDR-2014~\cite{froehlich2014creating} (top) and five held-out
    EventHDR sequences~\cite{zou2024eventhdr} (bottom). The 5$\times$
    models remain relatively stable over their short output horizons.
    For the 20$\times$ models, reconstruction quality degrades gradually
    as the temporal distance from the reference increases, with more
    noticeable quality loss at larger temporal offsets.
    }
    \label{fig:supp_perframe}
\end{figure}

We further analyze frame-wise reconstruction quality over the temporal
horizon to study how HIVE-Diff behaves as the output moves farther from
the single LDR reference. \cref{fig:supp_perframe} shows the trajectories
of SSI-PSNR~\cite{weigert2018content,krull2020probabilistic},
SSIM~\cite{wang2004image}, LPIPS~\cite{zhang2018unreasonable}, and
CIEDE2000~\cite{luo2001development} on the synthetic HdM-HDR-2014
benchmark~\cite{froehlich2014creating} and the held-out EventHDR
sequences~\cite{zou2024eventhdr}.

On the synthetic benchmark, the 5$\times$ model maintains relatively
stable quality across all five output frames. The 20$\times$ model is
also stable near the reference, followed by gradual degradation over the
longer horizon. SSI-PSNR decreases from 26.04 at $t{=}0$ to 24.15 at
$t{=}19$, while SSIM decreases from 0.752 to 0.702 and LPIPS increases
from 0.217 to 0.272. CIEDE2000 similarly increases from 7.52 to 9.48.
The smooth evolution of these metrics indicates a gradual loss of
frame-wise fidelity rather than an abrupt failure at longer temporal
offsets.

A similar overall trend is observed on the held-out EventHDR sequences.
The 5$\times$ model remains stable over its short horizon, while the
20$\times$ model exhibits gradual degradation as the temporal offset
increases. SSI-PSNR decreases from 23.15 at $t{=}0$ to 21.76 at
$t{=}19$, SSIM decreases from 0.835 to 0.793, and LPIPS increases from
0.204 to 0.245. CIEDE2000 exhibits larger frame-dependent fluctuations,
but increases overall from 11.19 to 11.82. These results show that
HIVE-Diff maintains relatively stable reconstruction quality over short
and intermediate temporal horizons, while longer single-reference
expansion becomes progressively more challenging.

\FloatBarrier

\section{Effect of Previous-Latent Conditioning}
\label{sec:supp_previous_latent}

\begin{figure}[H]
    \centering
    \includegraphics[width=0.56\textwidth]
    {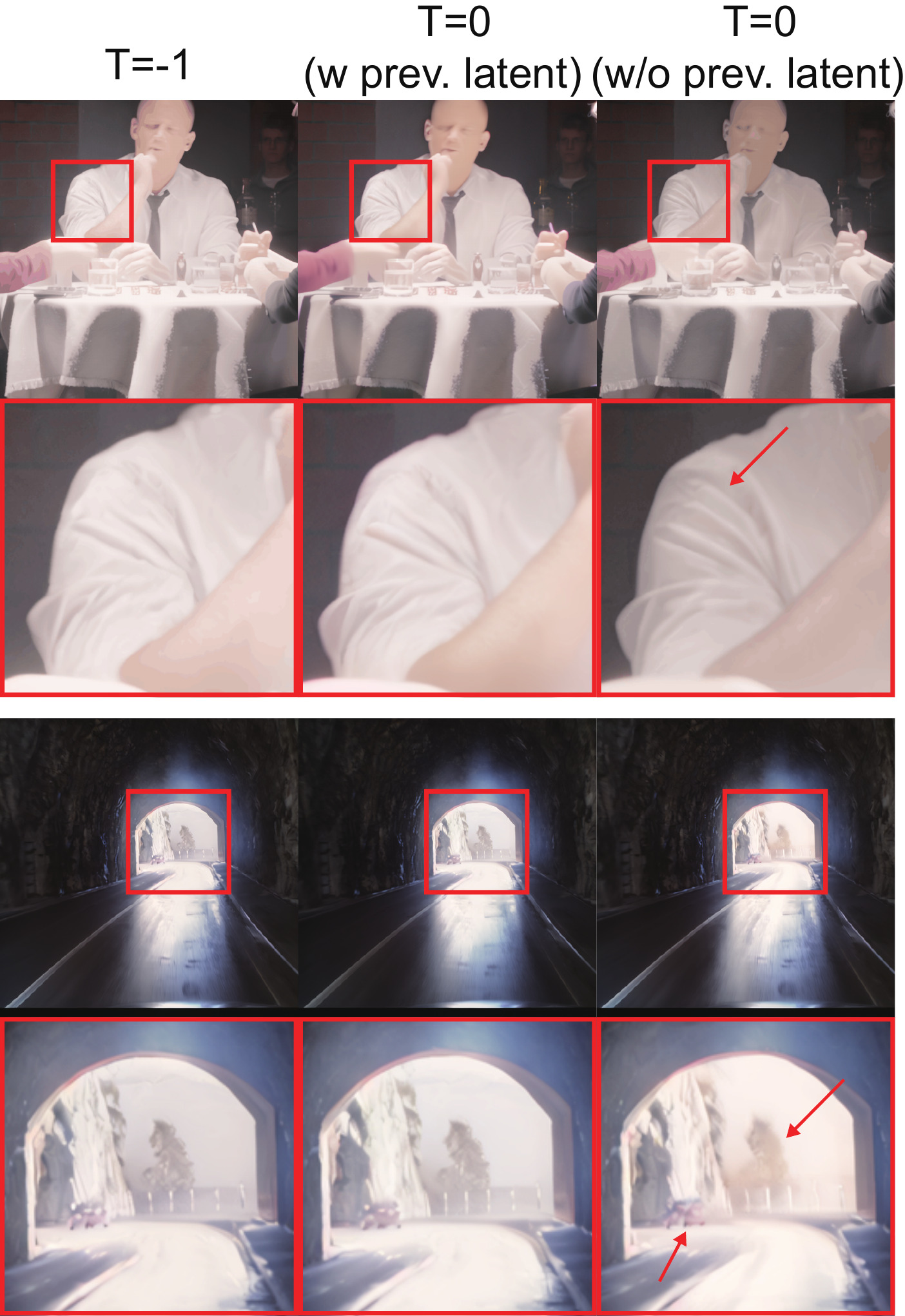}
    \caption{
    Effect of previous-latent conditioning in the 5$\times$ setting on
    (top) synthetic HdM-HDR-2014~\cite{froehlich2014creating} and
    (bottom) DSEC~\cite{Gehrig21ral}. The left column shows the last
    frame of the previous clip ($t{=}-1$), while the middle and right
    columns show the first frame of the next clip ($t{=}0$) with and
    without previous-latent conditioning, respectively. Frame indices
    are defined relative to the clip boundary. Previous-latent
    conditioning improves structural and appearance continuity, whereas
    independent generation introduces visible discontinuities highlighted
    by the red boxes and arrows.
    }
    \label{fig:supp_previous_latent}
\end{figure}

We further examine previous-latent conditioning in the 5$\times$
setting to evaluate continuity across independently generated clips.
As shown in \cref{fig:supp_previous_latent}, we compare the last frame
of the previous clip ($t{=}-1$) with the first frame of the subsequent
clip ($t{=}0$), generated with and without conditioning on the previously
denoised latent. Here, the frame indices are defined relative to the
clip boundary. Without previous-latent conditioning, independently
generated clips can exhibit visible discontinuities in local structure,
brightness, and appearance. Conditioning on the previous latent
substantially reduces these discontinuities and produces a smoother
cross-clip transition.

Previous-latent conditioning can be viewed as a short-term memory that
promotes local consistency between adjacent clips. However, it does not
explicitly constrain the global appearance of the entire sequence.
Consequently, small brightness and color deviations may accumulate
during long autoregressive generation, resulting in residual flickering
even when individual clip boundaries are locally smooth. Similar
long-horizon consistency issues have been studied in recent long-video
diffusion methods, where short-term inter-clip memory is complemented by
long-term appearance preservation~\cite{henschel2025streamingt2v}.

Importantly, subsequent clips may also receive newly observed LDR frames
and event measurements. Therefore, the HDR content of a new clip is
expected to update as the scene and observations evolve, and not all
cross-clip appearance changes should be suppressed. The key challenge is
to distinguish observation-driven content updates from undesired
generation drift. A promising extension of HIVE-Diff is thus to combine
the previous latent with a dynamically updated appearance memory derived
from the available LDR observations, rather than relying on a fixed
appearance anchor. Overlapping-window generation with latent, feature,
or photometric consistency could further suppress residual flickering
while allowing new LDR and event evidence to modify the reconstructed
HDR content. Long-video diffusion techniques based on temporal memory
and windowed generation~\cite{henschel2025streamingt2v,qiu2024freenoise}
provide useful directions for this extension.

\FloatBarrier

\clearpage
\section{Additional Qualitative Results}
\label{sec:supp_additional_qualitative}

We provide additional qualitative results on the synthetic
HdM-HDR-2014 benchmark~\cite{froehlich2014creating}, held-out
EventHDR sequences~\cite{zou2024eventhdr}, and real-world datasets
without HDR ground truth. In addition to the reference-aligned results
presented in the main paper, these examples show HIVE-Diff outputs at
multiple timestamps over the 20$\times$ reconstruction horizon.

\paragraph{Synthetic data.}

\begin{figure}[H]
    \centering
    \includegraphics[width=\textwidth]{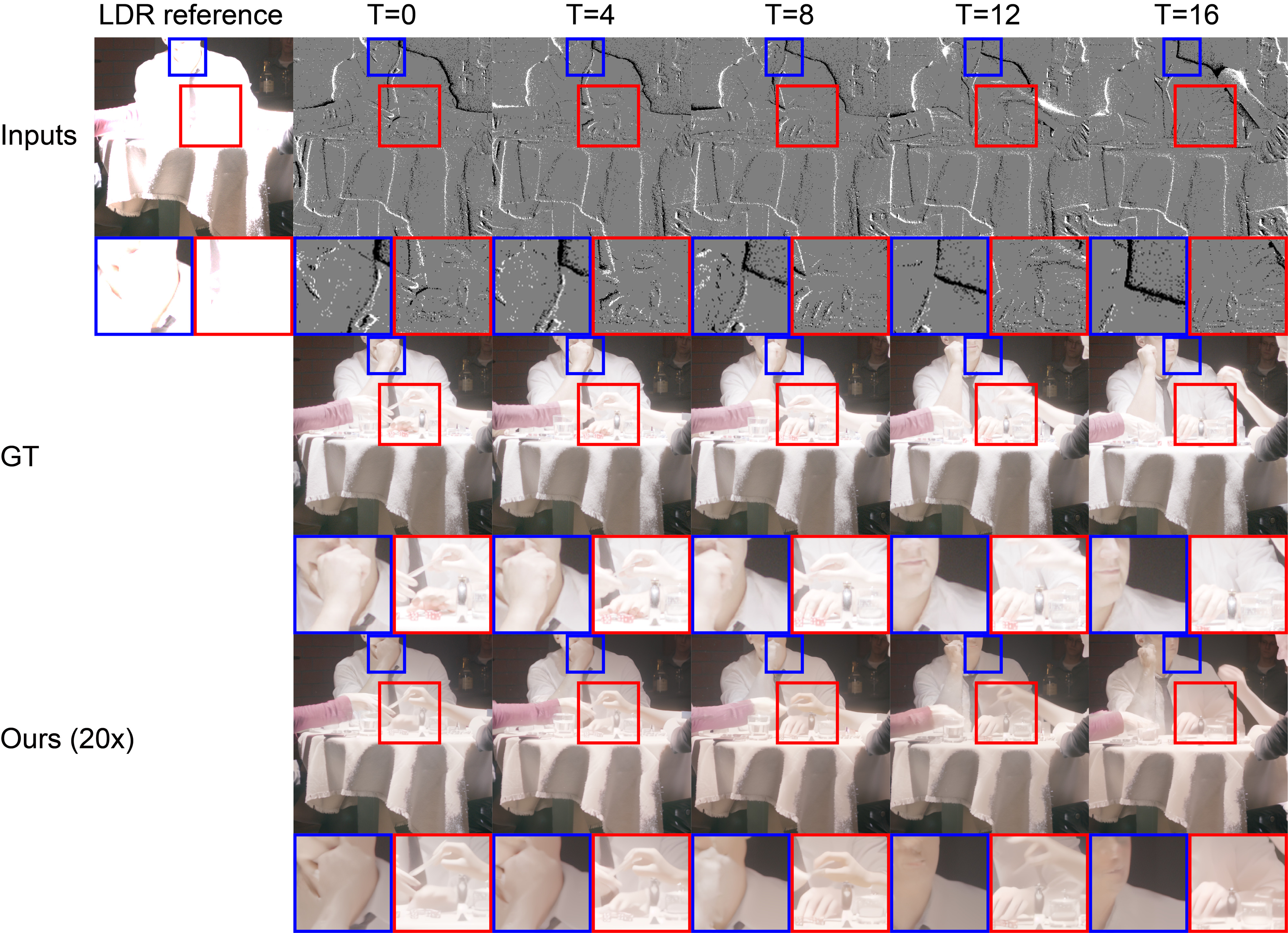}
    \caption{
    Additional qualitative results on HdM-HDR-2014
    \cite{froehlich2014creating}. The first row shows the LDR reference
    and event observations at selected timestamps. The second and third
    groups show the corresponding HDR ground truth and HIVE-Diff
    20$\times$ reconstructions, respectively. Enlarged regions highlight
    local reconstruction details across the temporal horizon.
    }
    \label{fig:supp_synthetic_more}
\end{figure}

\cref{fig:supp_synthetic_more} shows an additional example on
HdM-HDR-2014~\cite{froehlich2014creating}, where HDR ground truth is
available for direct comparison. Starting from a single over-exposed
LDR reference and the corresponding event stream, HIVE-Diff reconstructs
the scene over a long temporal horizon. The selected frames show that
the model preserves the overall scene structure and recovers content in
saturated regions across time. The enlarged regions further illustrate
the reconstruction of local appearance and motion at different temporal
offsets.

\FloatBarrier

\clearpage
\paragraph{EventHDR.}

\begin{figure}[H]
    \centering
    \includegraphics[width=\textwidth]{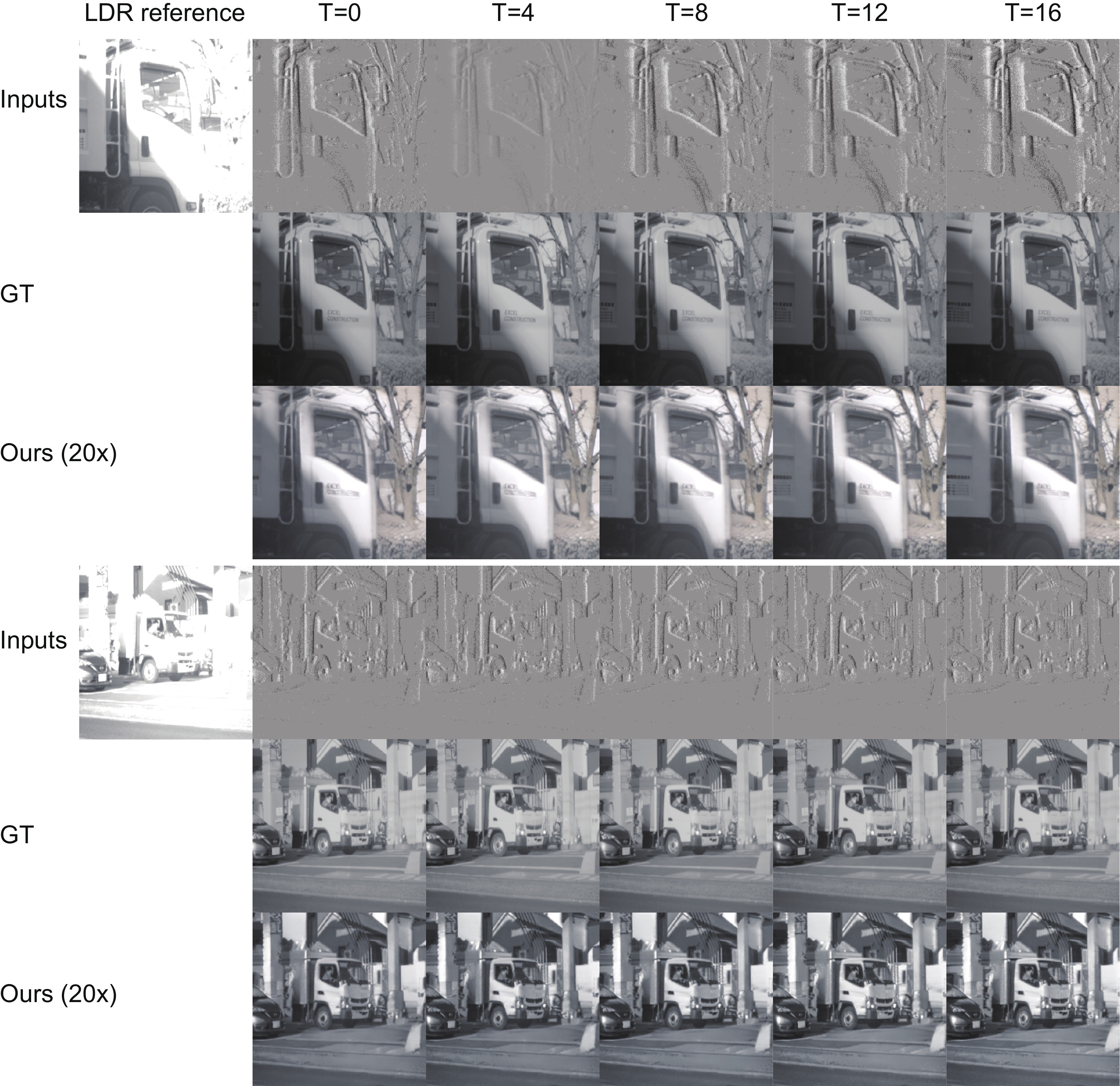}
    \caption{
    Additional qualitative results on two held-out EventHDR
    sequences~\cite{zou2024eventhdr}. For each example, the input
    consists of one LDR reference and event observations at selected
    timestamps. HDR ground truth and the corresponding HIVE-Diff
    20$\times$ reconstructions are shown for direct comparison.
    }
    \label{fig:supp_eventhdr_more}
\end{figure}

We next provide additional results on held-out EventHDR
sequences~\cite{zou2024eventhdr} in \cref{fig:supp_eventhdr_more}.
Despite using only the first LDR frame as the appearance and intensity
reference, HIVE-Diff follows the temporal evolution indicated by the
real event stream and reconstructs subsequent HDR frames over the
20$\times$ horizon. The reconstructed scene content remains visually
close to the available HDR ground truth across the selected timestamps.

\FloatBarrier

\clearpage
\paragraph{Real-world data without HDR ground truth.}

\begin{figure}[H]
    \centering
    \includegraphics[width=\textwidth]{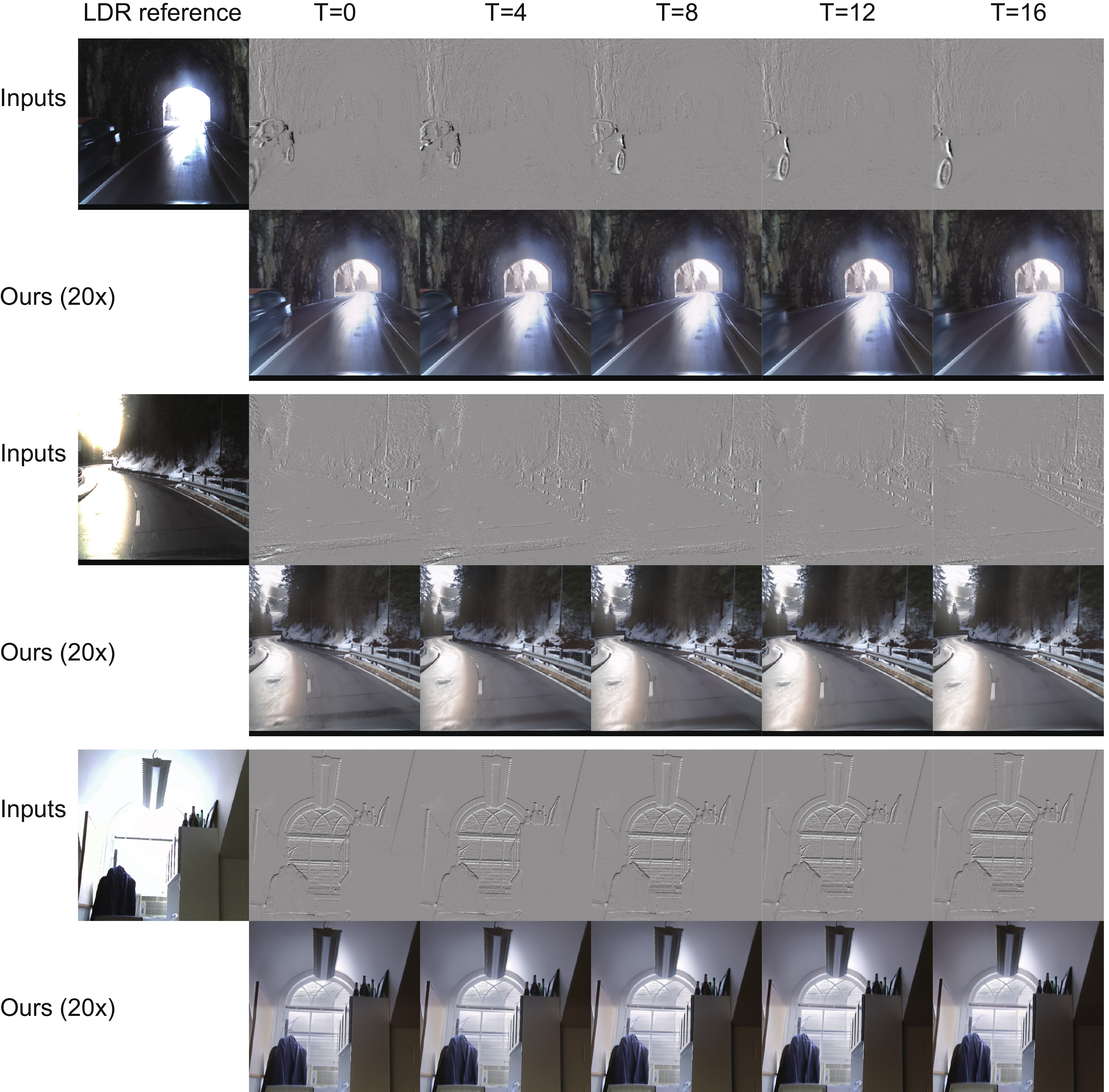}
    \caption{
    Additional qualitative results on real-world data without HDR ground
    truth. The top and middle examples are from DSEC~\cite{Gehrig21ral},
    while the bottom example is from our self-collected data. Each
    example shows one LDR reference, event observations at selected
    timestamps, and the corresponding HIVE-Diff 20$\times$
    reconstructions.
    }
    \label{fig:supp_real_more}
\end{figure}

Finally, \cref{fig:supp_real_more} presents additional results on two
DSEC sequences~\cite{Gehrig21ral} and one sequence from our self-collected
event-frame data, none of which provides HDR ground truth. Across these
scenes, HIVE-Diff reconstructs temporally evolving scene content from a
single LDR reference and the subsequent event stream while recovering
structures in severely over-exposed regions. These examples further
demonstrate the applicability of single-reference HDR temporal expansion
to real event-camera data. As discussed in
\cref{sec:supp_previous_latent}, mild brightness or color drift may still
appear over long sequences.

\FloatBarrier

\section{Runtime and Computational Cost}
\label{sec:supp_runtime}

Tab.~5 of the main paper reports model size, inference time per output
frame, computational cost, and peak GPU memory of HIVE-Diff and the
event-based HDR baselines HDRev~\cite{yang2023learning},
NeurImg-HDR~\cite{han2023hybrid}, EventHDR~\cite{zou2024eventhdr}, and
HDRev-Diff~\cite{yang2025event}, profiled at $512\times512$ resolution on
a single NVIDIA L40S GPU. HIVE-Diff uses the 5$\times$ model with 25-step
Euler sampling, FP16 precision, and classifier-free guidance (CFG),
without temporal chunking; HDRev-Diff uses 50-step DDIM sampling with
CFG. The two diffusion methods have similar computational cost per output
frame (120.6 vs.\ 120.4 TFLOPs), while HIVE-Diff has more parameters and
higher peak memory (9.2 vs.\ 6.7 GiB). Because the two methods use
different numerical precisions, samplers, and numbers of denoising steps,
the 2.3$\times$ runtime difference should be read under these inference
settings rather than as an intrinsic architectural speedup.

Further acceleration could be achieved through two complementary
directions. First, recent few-step video diffusion distillation methods
can compress a multi-step denoising trajectory into only a few sampling
steps~\cite{ding2025dollar}, suggesting that the current 25-step HIVE-Diff
sampler could potentially be distilled to substantially fewer steps.
Second, training-free feature caching can reduce the computation within
each remaining denoising step by reusing redundant intermediate features
across adjacent timesteps~\cite{ma2024deepcache}. Since HIVE-Diff also
uses CFG, CFG-aware caching methods such as FasterCache
\cite{lyu2025fastercache} are particularly relevant because they further
exploit redundancy between the conditional and unconditional branches.
Combining few-step distillation with feature/CFG caching is therefore a
promising direction for reducing HIVE-Diff inference latency, although
the achievable speedup and its effect on HDR reconstruction fidelity and
temporal consistency require dedicated evaluation.

\FloatBarrier

\end{document}